\documentclass[a4paper,preprint,superscriptaddress,showpacs,preprintnumbers,nofootinbib]{revtex4}
\usepackage{graphicx}
\usepackage{epstopdf}
\usepackage{dcolumn}
\usepackage{bm}

\newcommand{\nn}{\nonumber}

\newcommand{\be}{\begin{eqnarray}}
\newcommand{\ee}{\end{eqnarray}}
\catcode`\@=11
\def\lsim{\mathrel{\mathpalette\@versim<}}
\def\gsim{\mathrel{\mathpalette\@versim>}}
\def\@versim#1#2{\vcenter{\offinterlineskip
\ialign{$\m@th#1\hfil##\hfil$\crcr#2\crcr\sim\crcr } }}
\catcode`\@=12

\def\thefootnote{\fnsymbol{footnote}}

\begin{document}

\title{Two-loop radiative seesaw with multicomponent dark matter \\explaining the possible $\gamma$ excess in the Higgs boson decay \\and at the Fermi LAT}

\author{Mayumi \surname{Aoki}}
\email{mayumi@hep.s.kanazawa-u.ac.jp}
\affiliation{Institute for Theoretical Physics, Kanazawa University, Kanazawa 920-1192, Japan}
\affiliation{Max-Planck-Institut f\"ur Kernphysik, Saupfercheckweg 1, 69117 Heidelberg, Germany}

\author{Jisuke \surname{Kubo}}
\email{jik@hep.s.kanazawa-u.ac.jp}
\affiliation{Institute for Theoretical Physics, Kanazawa University, Kanazawa 920-1192, Japan}

\author{Hiroshi \surname{Takano}}
\email{takano@hep.s.kanazawa-u.ac.jp}
\affiliation{Institute for Theoretical Physics, Kanazawa University, Kanazawa 920-1192, Japan}

\preprint{KANAZAWA-13-02}

\pacs{95.35.+d, 95.85.Pw, 11.30.Er}
\keywords{Dark Matter,  Neutrino mass, Higgs, Fermi Lat}

\begin{abstract}
A non-supersymmetric model of a two-loop radiative seesaw
is proposed.
The model contains, in addition to the standard model  (SM) Higgs boson, 
an inert $SU(2)_L$ doublet scalar $\eta$ and two inert 
singlet scalars $\phi$ and $\chi$.
The lepton number is softly broken
by a dimension-two operator, and
the tree-level mass of the left-handed neutrino is forbidden by 
$Z_2 \times Z_2'$ (or $D_{2N}$) ,
which predicts the existence of two or three dark matter particles.
The scalar sector is minimal; none of the scalar fields can be suppressed
for the radiative seesaw mechanism to work.
There are by-products:
The SM Higgs boson decay into two $\gamma$'s is slightly enhanced 
by $\eta^+$ (the charged component of $\eta$) circulating 
in   one-loop diagrams for $h\to\gamma\gamma$.
The $135$ GeV $\gamma$-ray line
observed at the  Fermi LAT can be also explained
by the annihilation of $\chi$ dark matter.
We employ a mechanism of temperature-dependent
annihilation cross section to suppress the continuum $\gamma$ rays
and the production of antiprotons.
The explanation can survive even down to  the XENON1T sensitivity limit.

 \end{abstract}
\setcounter{footnote}{0}
\def\thefootnote{\arabic{footnote}}
\maketitle

\section{Introduction}

Why the neutrino masses are small
is a long-standing mystery.
The seesaw mechanism 
\cite{Minkowski:1977sc,seesaw,Mohapatra:1979ia}
is an approach to provide an answer to it.
The traditional seesaw mechanism
indicates the existence of a super-high-scale
physics beyond the standard model (SM).
Another way  to confront this  problem
is to generate the neutrino masses radiatively 
\cite{Zee:1980ai,Zee:1985id,Babu:1988ki}.
Many models  have been  proposed on 
the basis of the radiative generation of the neutrino masses,
and the idea of the radiative seesaw mechanism
\cite{Krauss:2002px,Ma:2006km}
is along the line of this idea:
Right-handed neutrinos are introduced, but
the Dirac masses are forbidden by a discrete symmetry.
This discrete symmetry can be an origin
of stable  dark matter (DM) particles
\cite{Krauss:2002px,Ma:2006km,Cheung:2004xm,Kubo:2006yx,Aoki:2008av}
 in the Universe.

To produce the neutrino masses, the lepton number L
has to be violated.  In  most of the models the lepton number L is
violated softly by dimension-three operators,  Majorana masses
or scalar trilinear couplings, and the number of the loops ranges
from one to three (see Refs.~\cite{Bonnet:2012kz,Farzan:2012ev} for different models).
For the radiative seesaw mechanism, the number of loops $\ell$ means
a scaling down of  
$(1+3\ell)$  orders of magnitude for  the right-handed neutrino mass
(see also the discussions of Ref.~\cite{Kanemura:2010bq});
$(k/16 \pi^2)^\ell \simeq (k /0.1) \times 6.3\times 10^{-4},~
(k /0.1)^2 \times 4.0\times 10^{-7},~
(k /0.1)^3 \times 2.5\times 10^{-10}$ for $\ell=1,~2$ and $3$, respectively,
where $k$ is a generic coupling.
 Since the Majorana mass of the right-handed neutrino for the tree-level seesaw is 
$O(10^{10})$ GeV, we may obtain a Majorana mass of $O(1)$ TeV naturally
in  two-loop radiative seesaw models 
\cite{Kanemura:2011jj,Law:2012mj,Guo:2012ne}.

 In this paper we propose a radiative seesaw model, in which
 the lepton number is
 softly broken by a dimension-two operator,
 and the neutrino masses are generated
 at the two-loop level. The discrete symmetry is 
 $Z_2\times Z'_2$ (or $D_{2N}$ with $N=2,3,\dots$
  \footnote{$D_{2N}$ (the dihedral group of order $2N$)
is larger than $Z_2\times Z'_2$.
However, we use only the one-dimensional 
representations of $D_{2N}$ so that
 $D_{2N}$ acts as $Z_2\times Z'_2$.}).
 Therefore, two or three DM particles can exist in this model
 \footnote{A multicomponent DM system has been considered recently in
 Refs.~\cite{D'Eramo:2010ep,Belanger:2011ww,Belanger:2012vp,Aoki:2012ub}; 
 see also the references therein.}, which is a slight extension
 of the Ma model \cite{Ma:2006km}.
Obviously,  radiative generation of the neutrino masses
means an extension of the SM Higgs sector,
which may have impacts on the existing experiments.
In our model, we have a set of 
an inert doublet scalar $\eta$ and two
singlet scalars. This set is minimal in the sense that
the  radiative neutrino mass generation does not work
 if one of them is suppressed.
So, none of the extra scalar fields is {\it ad hoc} introduced.

Another motivation to extend the original Ma model is the following: 
If a neutral component of $\eta$ in the Ma model should be 
a realistic DM particle,  its mass should be between 60 and 80 GeV
or larger than 500 GeV 
\cite{LopezHonorez:2006gr,Dolle:2009fn}
(unless one allows  very fine tuning of parameters \cite{LopezHonorez:2010tb}).
In Ref.~\cite{Aoki:2012ub}  we have slightly 
modified the Ma
model  such that there exist more than two stable DM particles
and have found that the $\eta$ DM mass 
is then allowed to  lie in a much wider range. 
But two scalars have been {\it ad hoc} added.

The existence of  additional scalar doublets
can change the decay rates of the SM Higgs boson $h$.
The results of the LHC indicate a slight
excess of $h\to \gamma\gamma$ \cite{:2012gk,:2012gu,:2012goa}, 
which in fact could be explained by an additional
inert  doublet circulating in 
one loop \cite{Gunion:1989we,Akeroyd:2007yh,Arhrib:2012ia,Swiezewska:2012eh}.
There is yet another excess of $\gamma$ 
at the Fermi Large Area Telescope (LAT) 
\cite{Atwood:2009ez,Abdo:2010nc,Ackermann:2012qk,Fermi-LAT}.
There are analyses  
\cite{Bringmann:2012vr,Weniger:2012tx,Tempel:2012ey,Rajaraman:2012db,Su:2012ft,Su:2012zg} 
that indicate
a monochromatic $\gamma$-ray line
of $135$ GeV
in  the Fermi data.
It has been  reported \cite{Biswas:2013nn} that a two-component
DM system consisting of  an inert  doublet scalar
and a scalar can explain the monochromatic $\gamma$-ray  line
at the Fermi LAT. 
Several models with  a two-component DM 
have also been considered in Refs.~\cite{D'Eramo:2012rr,Gu:2013iy} to explain the monochromatic Fermi LAT $\gamma$-ray line.
Therefore it is natural to wonder
 whether our two-loop radiative seesaw model can explain 
the $\gamma$ excess in the Higgs boson decay as well as
in the Fermi data
\footnote{
Recently, it has been argued that the monochromatic $\gamma$-ray line can be explained basically by the same one-loop contribution as for $h \to \gamma\gamma$ \cite{Biswas:2013nn,Baek:2012ub,Bai:2012nv,
Cline:2012nw, Wang:2012ts,Fan:2013qn}. }.
We  find that this is in fact possible if we accept that 
certain scalar couplings are large at the border
of perturbation theory, where
to  suppress sufficiently the  continuum
$\gamma$'s and the production of antiprotons,
we employ a mechanism of temperature-dependent
annihilation cross section \cite{Griest:1990kh,Tulin:2012uq}.

 \section{The Model}

\begin{table}
\caption{\footnotesize{The matter content of the model and
the corresponding quantum numbers. $Z_{2}
\times  Z'_2$ is  the unbroken discrete symmetry,
while the lepton number L is softly broken by the $\phi$ mass.
$D_{2N}~(N=2,3,\dots)$ is the dihedral group
of order $2N$.}}
\begin{center}
\begin{tabular}{|c|c|c|c|c|c|c||c|} 
\hline
Field & Statistics & $SU(2)_L$ & $U(1)_Y$  &L & $Z_2$ &$Z'_2$ & $D_{2N}$	\\ \hline
$L=(\nu_{L},l_L)$	& F& $2$ 	& $-1/2$ & $1$ & $+$ & $+$ & ${\bf 1}$\\ \hline
$l^c_R$ 		& F	& $1$ 	& $1$& $-1$ & $+$ & $+$ &${\bf 1}$	\\ \hline
$N^c_R$		& F		& $1$& $0$ & $0$ 	& $-$ & $+$ &${\bf 1''}$	\\ \hline
$H=(H^+,H^0)$ 	& B& $2$ 	& $1/2$ & $0$ & $+$ & $+$ &${\bf 1}$\\ \hline
$\eta=(\eta^+,\eta^0) 	$ 	& B& $2$ 	& $1/2$ & $-1$ & 
$-$ & $+$ &${\bf 1''}$	\\ \hline
$ \chi $     	& B  & $1$    & $0$  & $0$ & $+$ & $-$	 &${\bf 1'}$\\ \hline
$ \phi $   	& B  & $1$    & $0$  & $1$ & $-$ & $-$	 &${\bf 1'''}$\\ \hline
\end{tabular}
\end{center}
\end{table}
The matter content of the model is shown in Table I.
The new fields are (in addition to the right-handed neutrino $N_R^c$) 
the $SU(2)_L$ doublet scalar $\eta$ (L$=-1$) and  singlet scalars 
$\chi$ (L$=0$) and $\phi$ (L$=1$),
where L is the lepton number.
 Note that the L of $N_R^c$ is zero and
 that four different representations of $Z_2\times Z'_2$
 are exactly the singlets of the dihedral group of order
 $2N$, $D_{2N}~(N=2,3,\dots)$.
The $Z_2 \times Z'_2 \times\mbox{L}$--invariant 
(or $D_{2N}\times\mbox{L}$ --invariant) Yukawa 
couplings of the lepton sector can be described by
\begin{equation}
\mathcal{L}_Y 
= Y^e_{ij} H^\dag L_i  l_{Rj}^c
 + Y^{\nu}_{ik}L_i \epsilon \eta N_{Rk}^c
 + h.c. ~,
 \label{LY}
\end{equation}
with the Majorana mass term of the right-handed neutrinos 
$N_{Rk}^c~(k=1,2,3)$
\begin{equation}
\mathcal{L}_\mathrm{Maj} = -\frac{1}{2} 
[\, M_{k} N_{Rk}^c N_{Rk}^c +h.c.\,]~.
\end{equation}
The most general form of the 
$Z_2 \times Z'_2\times$ L--invariant  
 scalar potential is given by
\begin{eqnarray}
V_{\lambda}&=&
\lambda_1 (H^\dag H)^2
 +\lambda_2 (\eta^\dag \eta)^2
 + \lambda_3 (H^\dag H)(\eta^\dag \eta)
+\lambda_4 (H^\dag \eta)(\eta^\dag H)
 \nonumber \\
 & &
 + \gamma_1 \chi^4
 +\gamma_2 (H^\dag H)\chi^2
 + \gamma_3 (\eta^\dag \eta)\chi^2
 + \gamma_4 |\phi|^4
+ \gamma_5 (H^\dag H)|\phi|^2\nn\\
  & &+ \gamma_6 (\eta^\dag \eta)|\phi|^2
+ \gamma_7 \chi^2|\phi|^2
+ \frac{\kappa}{2} [\,(H^\dag \eta)\chi \phi+h.c.\,]~.
 \label{potential}
\end{eqnarray}
Note that  the ``$\lambda_5$ term'', 
$(1/2)\lambda_5 (H^\dag \eta)^2$,  is forbidden by L.
The $Z_2\times Z'_2$--invariant 
mass term is
\be
V_{m}&=&m_1^2 H^\dag H + m_2^2 \eta^\dag \eta 
+ \frac{1}{2} m_3^2 \chi^2+ m_4^2 |\phi|^2
+ \frac{1}{2} m_5^2 [\, \phi^2+(\phi^*)^2\,]~,
\label{v2A}
\ee
where the last term in Eq.~(\ref{v2A}) breaks L softly.
This is the only $Z_2\times Z'_2$--invariant
mass term which can break L softly.
In the absence of this term, there will be no neutrino mass.
The charged CP-even and CP-odd scalars are defined as
\be
H &=& \left(\begin{array}{c}H^+ \\ (v_h+h+i G)/\sqrt{2}\\
\end{array}\right)~,~
\eta = \left(\begin{array}{c}\eta^+ 
\\ (\eta_{R}^0+i\eta_{I}^0)/\sqrt{2}\\
\end{array}\right)~,~\phi=(\phi_R+i \phi_I)/\sqrt{2}~.
\ee
The tree-level masses of the scalars are given by
\be
m_h^2 &=&2 \lambda_1 v_h^2~,
~m^2_{\eta^\pm}=m_2^2+\frac{1}{2}\lambda_3 v_h^2
~,~m^2_{\eta_R^0}=m^2_{\eta_I^0}
=m_2^2+\frac{1}{2}(\lambda_3+\lambda_4) v_h^2~,\nn\\
m^2_{\phi_R} &=& m_4^2+m_5^2+\gamma_5 v_h^2
~,~m^2_{\phi_I}=m_4^2-m_5^2+\gamma_5 v_h^2
~,~m^2_{\chi}=m_3^2+\gamma_2 v_h^2~.
\ee

A supersymmetric $SU(5)$ UV completion of the model in the spirit of Ref.
\cite{Ma:2007kt}
is possible, where the scalar masses are   protected from a large  correction
coming from the quadratic divergence.
 In this case, the soft breaking of the lepton number will appear
as a soft-supersymmetric-breaking B term. 
Another way 
to get rid of  the quadratic divergence without introducing supersymmetry
is given in Refs.
\cite{Bardeen:1995kv, Aoki:2012xs}, where  classical 
conformal symmetry is used.
In this treatment the quadratic divergence in the scalar
masses does nothing wrong, so there is no need to protect them.

\subsection{Stability of the vacuum}
The potential $V_{\lambda}$ is  bounded below  if
\be
& & \lambda_1~,~ \lambda_2~, ~\gamma_1~,~\gamma_4 >0~,
\label{cond1}\\~
& & \lambda_3 >  -\frac{2}{3}\sqrt{\lambda_1\lambda_2}~,~
~\lambda_3+\lambda_4  >  -\frac{2}{3}\sqrt{\lambda_1\lambda_2}~,
 \label{cond2}\\
 & &
~\gamma_2 >  -\frac{2}{3}\sqrt{\lambda_1\gamma_1}~,~
~\gamma_5 >  -\frac{2}{3}\sqrt{\lambda_1\gamma_4}~,~
~\gamma_3 >  -\frac{2}{3}\sqrt{\lambda_2\gamma_1}~,\\
&&
\gamma_6 >  -\frac{2}{3}\sqrt{\lambda_2\gamma_4}~,~
\gamma_7 >  -\frac{2}{3}\sqrt{\gamma_1\gamma_4}~,~\\
& &  \lambda_1+\lambda_2+
\gamma_1+ \gamma_4
-\frac{2}{3}\left(\sqrt{\lambda_1\lambda_2}
+\sqrt{\lambda_1\gamma_1}
+\sqrt{\lambda_1\gamma_4}\right.\nn\\
& &\left.+\sqrt{\lambda_2\gamma_1}
+\sqrt{\lambda_2\gamma_4}
+\sqrt{\gamma_1\gamma_4}\right)
> |\kappa| \label{cond3}
\ee
are  satisfied. 
The minimum of $V_{\lambda}$ is zero 
if the inequalities above
are satisfied.
The discrete symmetry $Z_2\times Z'_2$
(or $D_{2N}$) is unbroken
if,  in addition to Eqs.~(\ref{cond1})--(\ref{cond3}), the inequalities
$ m_2^2 ,~m_3^2,~m_4^2$ and 
$m_4^2-|m_5|^2 > 0$ are satisfied.

The   inequalities of Eqs.~(\ref{cond1})--(\ref{cond3}) are sufficient
conditions, but not necessary ones. If we assume that
 Eq.~(\ref{cond1}) and $\gamma_2, \gamma_3,\gamma_5,
\gamma_6,\gamma_7 >0  $ are satisfied, for instance,  the inequalities of Eq.~(\ref{cond2}) are relaxed to
\be
& & \lambda_3~,~
 \lambda_3+\lambda_4  >  -2\sqrt{\lambda_1\lambda_2}~,
 \label{cond4}
 \ee
and Eq.~(\ref{cond3}) is relaxed to
\be
& &  \lambda_1+\lambda_2+
\gamma_1+ \sum_{i=2}^7\gamma_i
-2\sqrt{\lambda_1\lambda_2}
> |\kappa| ~.
\label{cond5}
\ee
Since $m_h=125$ GeV and $v_h=246$ GeV, 
the Higgs coupling $\lambda_1$ is fixed at $0.129$.
Then Eq.~(\ref{cond4}) implies that
\be
\lambda_3~,~ \lambda_3+\lambda_4 
 &> & -2.5 \sqrt{\lambda_2/4\pi}.
\label{cond6}
\ee

\subsection{Neutrino mass}
\begin{figure}
  \includegraphics[width=10cm]{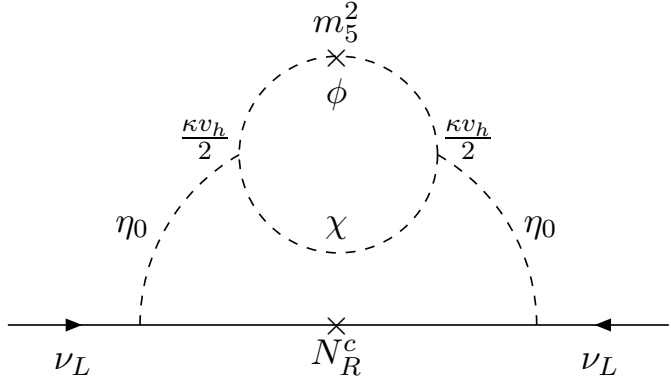}
\caption{\label{twoloop1}\footnotesize
Two-loop radiative neutrino mass.}
\end{figure}

The neutrino masses can be generated at the two-loop level
as shown in Fig.~\ref{twoloop1}.
The mechanism of the radiative generation is the following:
Because of  the soft breaking of the dimension-two operator
$\phi^2$,
the propagator between $\phi$ and $\phi$ 
can exist. This can generate a $\eta^0\eta^0$ mass term.
In the one-loop radiative seesaw model of Ref.~\cite{Ma:2006km}
this mass  is generated at the tree level through the 
``$\lambda_5$''
coupling.  So the effective  $\lambda_5^{\rm eff}$ is
\be
\lambda_5^{\rm eff} &=&-\frac{\kappa^2}{64\pi^2}
\left[\frac{m^2_{\phi_I}}{m^2_{\phi_I}
-m_\chi^2}\ln \frac{m^2_{\phi_I}}{m_\chi^2}-
\frac{m^2_{\phi_R}}{m^2_{\phi_R}
-m_\chi^2}\ln \frac{m^2_{\phi_R}}{m_\chi^2}
\right]~.\label{l5}
\ee
Therefore, $\lambda_5^{\rm eff}$ cannot be large within the framework
of perturbation theory ($\lambda_5^{\rm eff} \lsim O(10^{-2}))$.
For a set of parameter values
\be
m_{\chi} &=&  ~135~\mbox{GeV}~,~m_{\phi_R} = 
 ~300~\mbox{GeV}
~,~m_{\phi_I} =  ~700~\mbox{GeV}~,~\kappa=3
\label{aset}
\ee
for instance, we obtain $\lambda_5^{\rm eff}=-0.02$.
The two-loop neutrino mass matrix is calculated to be
\be
({\cal M}_\nu)_{ij} 
&=&\left(\frac{1}{16\pi^2}\right)^2
\frac{\kappa^2 v_h^2 }{8}
 \sum_k Y^\nu_{ik}Y^\nu_{jk}M_k ~
 \int_0^\infty dx \{~B_0(-x,m_\chi,m_{\phi_R})-B_0(-x,m_\chi,m_{\phi_I})~\}\nn\\
 & & \times \frac{x}{(x+m_\eta^2)^2(x+M_k^2)}~~
 \mbox{for}~m_\eta=m_{\eta_R^0}\simeq m_{\eta_I^0}~,
 \label{numass}
\ee
where
the function $B_0$ is the Passarino-Veltman function
\cite{Passarino:1978jh}
\be
\frac{i}{16\pi^2}B_0(p^2,m_1,m_2)=\int \frac{d^D k}{(2 \pi)^D}
\frac{1}{(k^2-m_1^2+i\epsilon)((k+p)^2-m_2^2+i\epsilon)}~.
\ee
For the set of parameters given in Eq.~(\ref{aset}) with
$M_k=1$ TeV, $m_\eta=150$ GeV and $v_h=246$ GeV we obtain
$({\cal M}_\nu)_{ij} \simeq -0.7 \times 10^7 \sum_k Y^\nu_{ik}Y^\nu_{jk}$ eV,
and so the neutrino mass will be $O(10^{-1})$ eV if 
$\sum_k Y^\nu_{ik}Y^\nu_{jk}=10^{-8}$.
If $m_{\eta} \ll M_k$, Eq.~(\ref{numass})  can be 
estimated as
\footnote{There is $O(1)$ correction to  the approximate formula Eq.~(\ref{numass1}),
which we have checked numerically.}
\be
({\cal M}_\nu)_{ij} &\sim &-\lambda_5^{\rm eff} v_h^2 \sum_k 
\frac{Y^\nu_{ik} Y^\nu_{jk} }{16 \pi^2 M_k}
\left(\ln\left(\frac{m_{\eta_R^0}}{M_k}\right)^2
+1\right)
\mbox{for}~m_{\eta} \ll M_k~.
\label{numass1}
\ee
Therefore, the scale of the light neutrino mass will be
\be
\frac{\kappa^2}{64\pi^2}\frac{1}{16\pi^2}\frac{m_D^2}{M}
\sim \left(\frac{\kappa}{0.1}\right)^2 10^{-7}\times \frac{m_D^2}{M}~,
\ee
where $m_D^2/M$ is the scale in the case of the tree-level  type-I seesaw.
This means that we can scale down the mass of the right-handed neutrino
by several orders of magnitude.
So, the right-handed neutrino masses of TeV or less are naturally expected
in this model.

\subsection{Constraints}

\noindent
{\bf 1}:  \underline{$\mu\to e ~\gamma$}\\
The constraint coming from
$\mu\to e \gamma$  is given by \cite{Ma:2001mr}
\begin{eqnarray}
&&B(\mu\rightarrow e\gamma)={3\alpha\over 64
\pi(G_Fm_{\eta^\pm}^2)^2}
\left| ~\sum_k Y^\nu_{\mu k}Y^\nu_{ek}F_2
\left({M_k^2\over m_{\eta^\pm}^2}\right)
\right|^2 \lsim 2.4\times10^{-12}~,\label{mug}\\
&&F_2(x)={1\over 6(1-x)^4}(1-6x+3x^2+2x^3-6x^2\ln x)~,\nonumber
\end{eqnarray}  
where the upper bound is taken from Ref.~\cite{Hayasaka:2010et}.
A similar, but slightly weaker bound for $\tau \to \mu (e) \gamma$ given
in Ref.~\cite{Hayasaka:2010et}
has to be satisfied, too.
Since $F_2(x) \sim 1/3x$ for $x\gg 1$, while  
$1/12 < F_2(x) < 1/6 $ for $0< x <1$, the constraint 
can be readily  satisfied if
$M_{k}\ll m_{\eta^\pm}$ or $M_{k}\gg m_{\eta^\pm}$.
If we assume that $M_1=M_2=M_3=1~\mbox{TeV}
\gg m_{\eta^\pm}$  in Eq.~(\ref{mug}),
the constraint of Eq.~(\ref{mug}) becomes
$B(\mu\rightarrow e\gamma)\simeq
10^{-7}\times |\sum_k Y^\nu_{\mu k}Y^\nu_{ek}|^2
  \lsim 2.4\times10^{-12}$.
Therefore, $|Y^\nu_{e k}Y^\nu_{\mu k}|^2\lsim O(10^{-5})$
can satisfy the constraint.\\

\noindent
{\bf 2}:  \underline{$g_\mu-2$}\\
The extra contribution  to the anomalous magnetic
moment of the muon, $a_\mu=(g_\mu-2)/2$, is given 
by \cite{Ma:2001mr} 
\begin{eqnarray}
\delta a_\mu &=& {\frac{m_\mu^2}{16 \pi^2 m_{\eta\pm}^2}}
\sum_k Y^\nu_{\mu k}Y^\nu_{\mu k}F_2
\left({M_k^2\over m_{\eta^\pm}^2}\right)~ .
\end{eqnarray}  
If we assume that 
$|\sum_k Y^\nu_{\mu k}Y^\nu_{\mu k} F_2
\left({M_k^2\over m_{\eta^\pm}^2}\right)|\simeq
|\sum_k Y^\nu_{\mu k}Y^\nu_{ek} F_2
\left({M_k^2\over m_{\eta^\pm}^2}\right)|$,
then we  obtain 
\be
|\delta a_\mu | &\simeq & 1.4\times 10^{-7} 
B(\mu\rightarrow e\gamma)^{1/2}~,
\ee
where the upper bound on $|\delta a_\mu|$ is given by $3.4 \times 10^{-11}$
\cite{Beringer:1900zz}. So, the constraint from $a_\mu$ has no significant 
influence, if the constraint of Eq.~(\ref{mug}) is satisfied.\\

\noindent
{\bf 3}: \underline{Electroweak precision}\\
The electroweak precision measurement requires
\cite{Barbieri:2006dq,Beringer:1900zz}
\be
\Delta T &\simeq & 1.08
\left(\frac{m_{\eta^\pm}-m_{\eta^0_R}}{v}\right)
\left(\frac{m_{\eta^\pm}-m_{\eta^0_I}}{v}\right)
=0.07\pm 0.08 \label{cnd4}
\ee
for $m_h=115.5$--$127$ GeV.
Therefore,  
$|m_{\eta^\pm}-m_{\eta^0_R}|,~
|m_{\eta^\pm}-m_{\eta^0_I}| \lsim 90~\mbox{GeV}$ 
is sufficient to meet the requirement.

\section{DM and restricting the parameter space}
\subsection{$h \to \gamma \gamma$}

\begin{figure}
  \includegraphics[width=12cm]{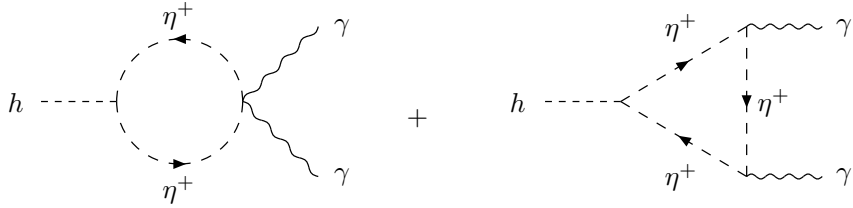}
\vspace{1cm}\caption{\label{hgg}\footnotesize
One-loop diagrams  for $h\to \gamma\gamma$.}
\end{figure}

\begin{figure}
  \includegraphics[width=10cm]{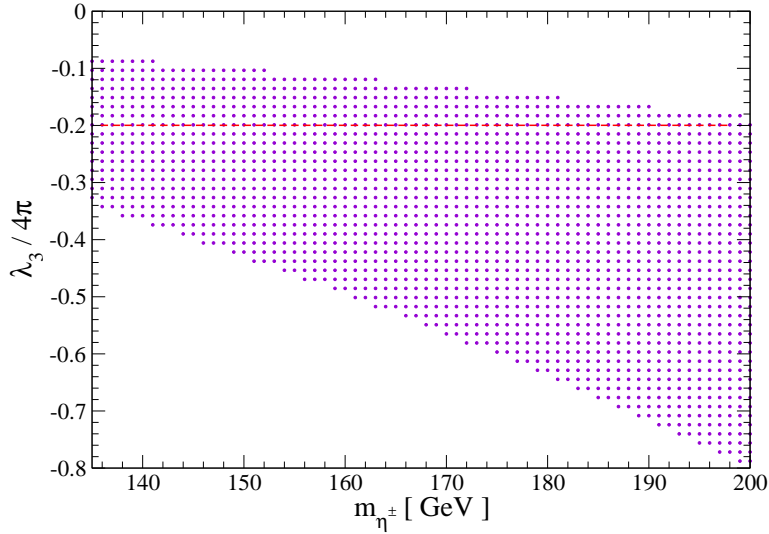}
\caption{\label{lambda3}\footnotesize
The area with $\Gamma/\Gamma^{\rm SM} 
=1.6\pm 0.4$ in the 
$m_{\eta^\pm}$--$\lambda_3$ plane.
The horizontal red line is the stability bound in given in Eq.~(\ref{cond6}).}
\end{figure}

Because of  the coupling
${\cal L}_{H^\dag H \eta^\dag \eta}=
-\lambda_3 (H^\dag H)(\eta^\dag\eta)
=-\lambda_3 v_h h \eta^+ \eta^- +\cdots$,
there are additional diagrams that contribute to 
the one-loop decay $h \to \gamma \gamma$,
which are shown in Fig.~\ref{hgg}.
Therefore, the decay width for two 
$\gamma$'s can be increased 
\cite{Gunion:1989we,Akeroyd:2007yh,Arhrib:2012ia,Swiezewska:2012eh}:
\be
\frac{\Gamma(\gamma \gamma)}
{\Gamma^{\rm SM}(\gamma \gamma)} &=&
\left[\frac{3 (3/2)^2 F_{1/2}(\tau_t)+F_1(\tau_W)
+2 \lambda_3 (m_W^2/g^2 m^2_{\eta^\pm})
F_0(\tau_{\eta^\pm})}{3 (3/2)^2 F_{1/2}(\tau_t)+F_1(\tau_W)}\right]^2~,\\
\tau_t &=& 4 m_t^2/m_h^2,~\tau_W = 4 m_W^2/m_h^2~,~
\tau_{\eta^\pm} = 4 m^2_{\eta^\pm}/m_h^2~,\nn\\
F_{1/2}(\tau) &=&2+3 \tau+3\tau (2-\tau )\arcsin^2(1/\sqrt{\tau})~,\nn\\
F_1(\tau) &=&-2\tau[1+(1-\tau)\arcsin^2(1/\sqrt{\tau})]~,\nn\\
F_0(\tau) &=&\tau[1-\tau\arcsin^2(1/\sqrt{\tau})]~,\nn
\ee
where $\arcsin^2(1/\sqrt{\tau})$ should be replaced
by $  (-1/4)[\ln \frac{1+
\sqrt{1+\tau}}{1-\sqrt{1-\tau}}-i \pi]^2$ 
for $\tau < 1$.
We obtain
\be
\frac{\Gamma(\gamma \gamma)}
{\Gamma^{\rm SM}(\gamma \gamma)} 
&\simeq &\left\{\begin{array}{ccc}
1.20 & 1.31  & 1.54 \\
1.17 & 1.26  & 1.45 \\
1.16 & 1.24  & 1.42 \\
1.05 & 1.12 & 1.22 \\
\end{array}\right.~\mbox{for}~
\lambda_3=-\{\begin{array}{ccc}
1 & 1.5 & 2.5 \\
\end{array}~
\mbox{and}~
m_{\eta\pm}=\left\{\begin{array}{c}
135 \\
145 \\
150 \\
200 \\
\end{array}\right.~\mbox{GeV}~.
\label{gg}
\ee
So, if the charged inert scalar $\eta^+$ is 
relatively light
and $\lambda_3$ is  negative and large, 
the observed excess $1.6\pm0.4$ 
in the CMS experiment \cite{:2012gu} can be explained.
(The best-fit signal strength for this mode in the ATLAS
experiment \cite{:2012gk,:2012goa}
is $\hat{\mu}= 1.9\pm 0.5$.)
In Fig.~\ref{lambda3} we show the area 
in the $m_{\eta^\pm}$--$\lambda_3$ plane
in which
$\Gamma(\gamma \gamma)/
\Gamma^{\rm SM}(\gamma \gamma)
=1.6\pm 0.4$ can be obtained
\footnote{The upper bound  $m_{\eta^\pm}
< 135$ GeV given in Ref. 
\cite{Swiezewska:2012eh} is obtained from 
$\Gamma(\gamma \gamma)/
\Gamma^{\rm SM}(\gamma \gamma)>1.3$ 
and   $\lambda_3/4\pi > -1.46/4\pi
\simeq -0.116$,
which is consistent with Eq.~(\ref{gg}).
Note that we use 
$2.0 \geq \Gamma(\gamma \gamma)/
\Gamma^{\rm SM}(\gamma \gamma) \geq1.2$ 
for Fig.~\ref{lambda3} 
along with the stability constraint [Eq.~(\ref{cond6})].}.
As we can see from Eq.~(\ref{cond2}),
a large negative $\lambda_3$ may endanger
the vacuum stability, because $\lambda_1$ is fixed at
$0.129$. Therefore, we assume that 
all the quartic scalar couplings except 
$\lambda_3$ and $\lambda_4$
are positive and use the second set of the inequality
conditions in Eqs. 
(\ref{cond4}) and (\ref{cond5}). 
Equation (\ref{cond6}) means that $\lambda_3 \gsim -2.5$
if $\lambda_2$ is at the border of perturbation theory.
Thus, the Higgs boson decay mode $h\to \gamma\gamma$ 
prefers the parameter
space:
\begin{enumerate}
\item
All the quartic scalar couplings except 
$\lambda_3$ and $\lambda_4$ are positive, and 
$\lambda_2$ is large.
\item
The constraints
\be
-0.2 \lsim \lambda_3/4 \pi  \lsim -0.08~\mbox{and}~m_{\eta^\pm}
\lsim 200~\mbox{GeV}~
\label{cond-r3}
\ee
have to be satisfied.
\end{enumerate}

\subsection{Direct detection of DM}
As we can see from Table I,
either $N_R^c$ or  $\eta$ can be a DM candidate.
Here we assume that $\eta_R^0$, the CP-even component of $\eta$,
is a DM particle and assume that $M_k \gg m_{\eta_R^0},~
m_{\eta_I^0} ,~m_{\eta^\pm}$ to
satisfy the $\mu\to e \gamma$ constraint [Eq.~(\ref{mug})].
The model can have three stable DM particles in principle,
but to simplify the situation we assume  a two-component DM system.
Another one is either $\chi$ or  $\phi$.
As we see from the potential [Eq.~(\ref{potential})], there is no
significant difference between $\chi$ and $\phi$
as DM. So we assume here that $\chi$ is the second DM
particle.

\begin{figure}
  \includegraphics[width=13cm]{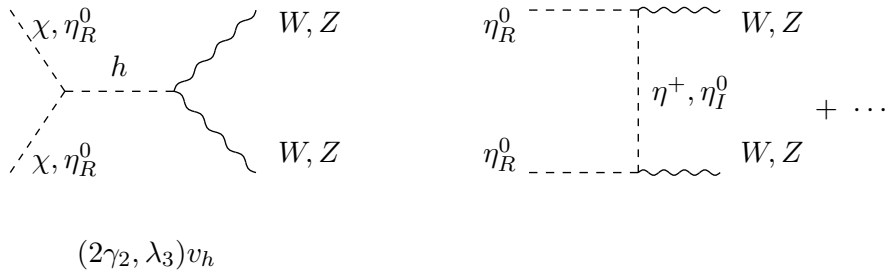}
\caption{\label{xxww}\footnotesize
Tree-level DM annihilations into  $W^+W^-$ and $Z Z$.}
\end{figure}
The spin-independent elastic cross sections off the nucleon,
$\sigma(\chi)$ and $\sigma(\eta^0_R)$,
are given by \cite{Barbieri:2006dq}:
\be
\sigma(\chi)
&=&\frac{1}{\pi} 
\left( \frac{\gamma_{2} \hat{f}
m_N }{m_{\chi} m_h^2} \right)^2
\left(\frac{m_N m_{\chi}}{m_N+m_{\chi}}
\right)^2,~
\sigma(\eta^0_R)
=\frac{1}{\pi} 
\left( \frac{\lambda_L \hat{f}
m_N/2 }{m_{\eta_R^0} m_h^2} \right)^2
\left(\frac{m_N m_{\eta_R^0}}{m_N+m_{\eta_R^0}}
\right)^2~,
\ee
where $\lambda_L=\lambda_3+\lambda_4$, $m_N$ is the nucleon mass, and
$\hat{f}\sim 0.3$ stems from the nucleonic matrix element 
\cite{Ellis:2000ds}.
The cross sections  have to satisfy
\be
\left(\frac{\sigma(\chi)}{\sigma_{\rm UB}(m_\chi)}\right)
\left(\frac{\Omega_\chi h^2}{\Omega_T h^2}\right)+
\left(\frac{\sigma(\eta_R^0)}{\sigma_{\rm UB}(m_{\eta_R^0})}\right)
\left(\frac{\Omega_\eta h^2}{\Omega_T h^2}\right) \lsim 1~,
\label{cond-omega}
\ee
where $\Omega_T h^2=0.116$ \cite{Hinshaw:2012fq}, and
 $\sigma_{\rm UB}(m)\simeq
3\times 10^{-45}~\mbox{cm}^2$ 
\cite{Aprile:2012nq}
 is the  XENON100 limit  for the DM mass of $135$ GeV.
So,
we  find
\be
|\gamma_2 | 
&\lsim & 0.035~ (0.003)~\mbox{if}~\Omega_\chi h^2\simeq 
\Omega_Th^2=0.116~\mbox{and}~
m_\chi=135~\mbox{GeV}~,
\label{cond-r2}\\
|\lambda_3+\lambda_4| 
&\lsim &0.069~ (0.006) ~\mbox{if}~\Omega_\eta h^2\simeq 
\Omega_Th^2=0.116 ~\mbox{and}~
m_{\eta^0_R}=135~\mbox{GeV}~,
\label{cond-r34}
\ee
where we would obtain the numbers in the parentheses when 
the XENON1T sensitivity  $\sigma\sim
2\times 10^{-47}~\mbox{cm}^2$ \cite{Aprile:2012zx} 
 has been reached.
Because of Eq.~(\ref{cond-r3}), we have to 
assume a large negative $\lambda_3$ and,
because of $m^2_{\eta^\pm}
-m^2_{\eta_R^0}=-\lambda_4 v_h^2/2$, 
 $\lambda_4$ has to be negative, too,
to ensure  $m_{\eta^\pm}> m_{\eta_R^0}$.
If $\lambda_4$ is negative and 
in addition  $\lambda_3$ is large and negative,
 the inequality of Eq.~(\ref{cond4}) may be violated, unless $\lambda_4$
is small and negative. This means $\lambda_L\simeq \lambda_3$,
implying that Eq.~(\ref{cond-r34}) cannot be satisfied.
Therefore, the relic density of $\eta_R^0$ has to
be small to satisfy  the constraint of Eq.~(\ref{cond-omega}).
This is welcome, because the annihilation cross section
of $\eta_R^0$ is large in general 
due to the gauge interactions shown in Fig.~\ref{xxww}
\cite{LopezHonorez:2006gr,Dolle:2009fn,LopezHonorez:2010tb}. 
Thus, the parameter
space   is further constrained:
\begin{enumerate}
\item
$\lambda_4$ has to be  negative and small
to ensure $m_{\eta^\pm} >
m_{\eta_R^0}$ and to satisfy the constraint of Eq.~(\ref{cond6}).
\item
Since $\lambda_3$ is assumed to be large and negative
[ see Eq.~(\ref{cond-r3}) ], the constraint of Eq.~(\ref{cond-omega}) can be satisfied 
only if $\Omega_\eta/\Omega_T \ll 1$. 
\item
To satisfy the XENON100 (1T) constraint, 
we have to impose $
|\gamma_2| \lsim  0.035 (0.003)$
for $m_\chi=135$ GeV.
\end{enumerate}

 \subsection{Relic densities of DM}
Since our parameter space has already been 
constrained to a certain amount,
we next calculate the relic density of DM, $\Omega_T=
\Omega_\chi+\Omega_\eta$.
To simplify the situation
we have been  assuming throughout  
that $\eta_R^0$ and $\chi$ are DM particles.
In this two-component DM system 
there are three different 
 thermally averaged cross sections
  \be
 \langle \sigma (\eta_R^0\eta_R^0;
\mbox{SM}) v \rangle
 & , & 
 \langle \sigma (\chi\chi;\mbox{SM}) v\rangle~,~
   \langle \sigma (\eta_R^0\eta_R^0;\chi\chi ) v\rangle ~
    \label{input1}
   \ee
that  are relevant for calculating
the DM relic density, where $\mbox{SM}$ stands for 
the SM particles\footnote{These  thermally averaged cross sections
 are tree-level ones and do not include
those  into two $\gamma$'s. Annihilations into two $\gamma$'s
will be separately calculated later on.}.
No semiannihilation  
 $\eta_R^0 \chi \to \phi_{R(I)}  \mbox{SM}$ is  allowed 
if  $m_{\phi_{R(I)}} > m_{\eta_R^0}+m_\chi$.
Then the evolution equation for $Y$, the number density
over the entropy density, can be written as
\cite{D'Eramo:2010ep,Belanger:2011ww,Belanger:2012vp,Aoki:2012ub}
  \be
\frac{d Y_{\eta_R^0}}{dx}
&=&
-0.264~ g_*^{1/2} \left[\frac{\mu M_{\rm PL}}{x^2} \right]
\Biggl\{
\langle\sigma (\eta_R^0\eta_R^0;\mbox{SM}) v\rangle
\left(  Y_{\eta_R^0}Y _{\eta_R^0}-\bar{Y}_{\eta_R^0}
\bar{Y}_{\eta_R^0}\right)
\nn\\ & &
+\langle\sigma (\eta_R^0\eta_R^0;\chi \chi )v\rangle
\!\!\left(  Y_{\eta_R^0} Y _{\eta_R^0}-\frac{Y_{\chi } Y_{\chi }}
{\bar{Y}_{\chi }\bar{Y}_{\chi }} 
\bar{Y}_{\eta_R^0}\bar{Y}_{\eta_R^0}
\right)
\Biggr\}
~,\label{boltz21}\\
\frac{d Y_{\chi }}{dx}
&=&
-0.264~ g_*^{1/2} 
\left[\frac{\mu M_{\rm PL}}{x^2} \right]
\Biggl\{
\langle\sigma (\chi \chi ;\mbox{SM}) v\rangle
\left(  Y_{\chi } Y _{\chi }-\bar{Y}_{\chi }\bar{Y}_{\chi }\right)
\nn\\ & &
-\langle\sigma (\eta_R^0\eta_R^0;\chi \chi )v\rangle
\!\!\left(  Y_{\eta_R^0} Y _{\eta_R^0}-\frac{Y_{\chi } Y_{\chi }}
{\bar{Y}_{\chi }\bar{Y}_{\chi }} \bar{Y}_{\eta_R^0}\bar{Y}_{\eta_R^0}
\right)
\Bigg\}
~,
\label{boltz22}
\ee
where $\bar{Y}$ is
 $Y$ in equilibrium,
 $x=\mu/T$, $1/\mu=1/m_{\eta_R^0}+1/m_{\chi}$, and $ T, 
M_{\rm PL} $ and $ g_*=90$ are 
the temperature,
the Planck mass and the total number of effective
degrees of freedom, respectively.

\begin{figure}
  \includegraphics[width=11cm]{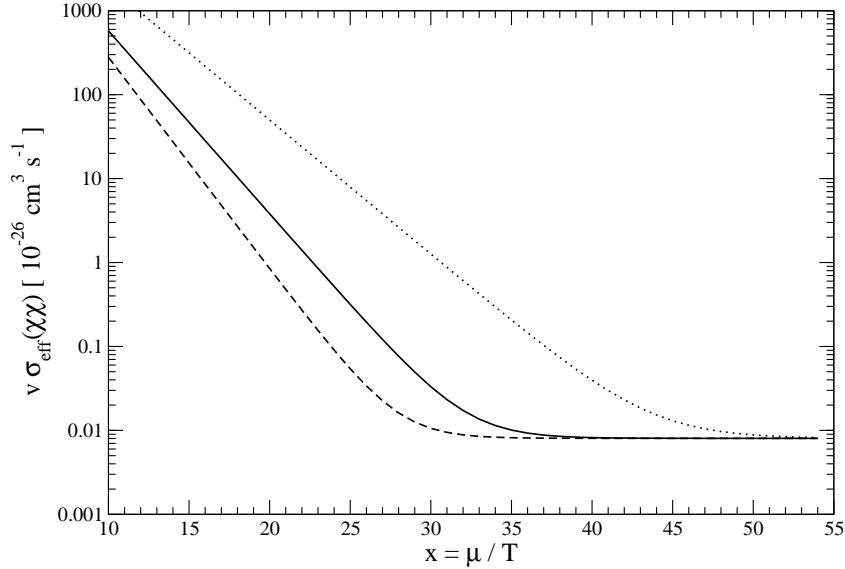}
\caption{\label{sigmaeff}\footnotesize
The effective annihilation cross section [ the expression in [ ] of Eq.~(\ref{s-eff}) ]
  as a function of $x=\mu/T$ for $m_{\eta_R^0}=148$ (dotted),
$153$ (solid), and $156$(dashed)  GeV with $m_\chi$ fixed at $135$ GeV.}
\end{figure}

Before we solve the evolution equations numerically,
we consider what we would expect.
As noticed, the relic density of $\eta_R^0$ will be very small
because of large $\lambda_3$ and gauge interactions (i.e.  large  
$\langle\sigma (\eta_R^0\eta_R^0;\mbox{SM}) v\rangle$), while
the annihilation cross section of $\chi$ into the SM
particles is  suppressed  because of Eq.~(\ref{cond-r2})
(i.e.  small  $\langle\sigma (\chi\chi;\mbox{SM}) v\rangle$).
That is, the DM conversion cross section
$\langle\sigma (\eta_R^0\eta_R^0;\chi\chi ) v\rangle $ 
and  the mass difference
$\Delta m_{\eta\chi}=m_{\eta_R^0}-m_\chi$ will play
an important role. Note that
the smaller $\Delta m_{\eta\chi}$ is, the larger is the effect
of the DM conversion  on $\Omega_\chi$.
To see this more explicitly, we assume that $\eta_R^0$ annihilates 
 very fast so that 
before and at the decoupling of $\chi$ the $\eta_R^0$
DM is in thermal equilibrium.
Then the expression in $\{~\}$ in the rhs of Eq.~(\ref{boltz22})
can be written as
\be
\left[\langle\sigma (\chi \chi ;\mbox{SM}) v\rangle
+\langle\sigma (\eta_R^0\eta_R^0;\chi \chi )v\rangle
\frac{m^3_{\eta_R^0}}{m_\chi^3}
\exp \left(2x\frac{m^2_\chi-m^2_{\eta_R^0}}{m_\chi m_{\eta_R^0}}\right) \right]
\left(  Y_{\chi } Y _{\chi }-\bar{Y}_{\chi }\bar{Y}_{\chi }\right)~,\label{s-eff}
\ee
which also appears in the coannihilation of DM with an unstable 
particle \cite{Griest:1990kh}
\footnote{The mechanism has been also used 
in the model of Refs.~\cite{Tulin:2012uq,Baek:2012ub}, 
explaining the monochromatic $\gamma$ at the Fermi LAT.
But the light charged scalar faces a problem in explaining the neutrino mass.}.
If $m^2_\chi-m^2_{\eta_R^0} <0$, 
the effective annihilation cross section
of $\chi$ is small at low temperature (large $x$),
while it is  large at high temperature (small $x$).
Because of the nontrivial interplay between 
$\gamma_2$ and $\Delta m_{\eta\chi}$, it may be possible
to obtain a correct relic density 
$\Omega_T h^2=0.1157\pm0.0023$
  \cite{Hinshaw:2012fq}.
  In Fig.~\ref{sigmaeff}  we show the effective
  annihilation cross section [ the expression in [ ] of Eq.~(\ref{s-eff}) ]
  as a function of $x=\mu/T$ for $m_{\eta_R^0}=148$ (dotted),
$153$ (solid), and $156$ (dashed)  GeV,
 where we have fixed the
 parameters as
\be
\lambda_3 &=&-1.26~,~
\lambda_4=-0.0205~,~
\gamma_3=11.3~,
\label{input2}\\
m_{\eta^\pm} &=& m_{\eta_I^0} =m_{\eta_R^0}
+4~\mbox{GeV}~,~
m_\chi =135~\mbox{GeV}~,~m_h=125~\mbox{GeV}~.
\label{input3}
\ee
 As we see from Fig.~\ref{sigmaeff}
the effective cross section  around the decoupling temperature $x\sim 20$
has a correct size and decreases drastically  at low temperature.
 The effective cross section is normalized to $10^{-26}~
\mbox{cm}^{3}\mbox{s}^{-1}$, because it is the size  to obtain the
observed relic density of DM.
The scalar couplings $\lambda_2,\gamma_1$ and 
$\gamma_4$ do not enter into the cross sections
[ Eq. (\ref{input1}) ], and $\gamma_5,
\gamma_6,\gamma_7$ and $\kappa$ 
are irrelevant because $\phi$  is much heavier than $\eta$ and $\chi$.

\begin{figure}
  \includegraphics[width=12cm]{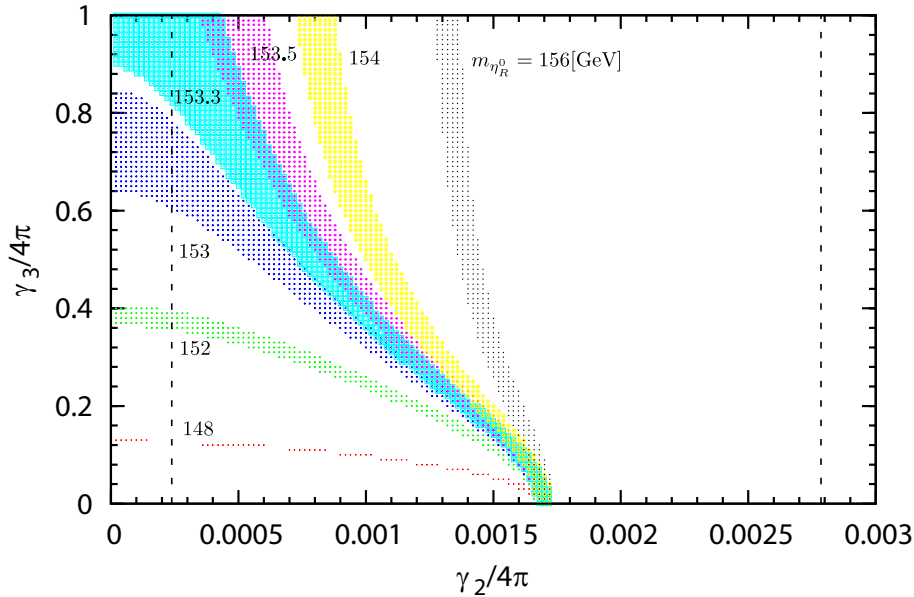}
\caption{\label{g2-g3}\footnotesize
The area with $\Omega_T h^2=
 0.1157\pm 0.0046~ (2\sigma)$ and $m_\chi=135$ 
 GeV in the $\gamma_2$--$\gamma_3$ plane
  for 
$m_{\eta_R^0}=148~(\mbox{red})$, $152~(\mbox{green})$, $153~(\mbox{blue})$, 
$153.3  ~ ( \mbox{cyan})$, $153.5  ~ (\mbox{purple})$, $154  ~ (\mbox{yellow})$  and $156 ~  (\mbox{black})$ GeV.
  The vertical (black dashed) lines are the upper bounds of $\gamma_2$ set by
  the XENON 100 (right) 
  and 1T (left) experiments [Eq.~(\ref{cond-r2}) ].}
\end{figure}

In Fig.~\ref{g2-g3} we  show the area in the $\gamma_2$--$\gamma_3$ plane
 in which the total relic density $\Omega_T h^2=
 0.1157\pm 0.0046 ~(2\sigma)$ with $m_\chi=135$ GeV can be obtained
  for 
$m_{\eta_R^0}=148~(\mbox{red})$, $152~(\mbox{green})$, $153~(\mbox{blue})$, 
$153.3  ~ ( \mbox{cyan})$, $153.5  ~ (\mbox{purple})$, $154  ~ (\mbox{yellow})$  and $156 ~  (\mbox{black})$ GeV.
  The right-handed neutrino masses,
$M_k$, are all $1$ TeV, and the Yukawa
couplings are chosen to yield
 $\sum_{ik}|Y^\nu_{ik}|^2=(10^{-4})^2$.
The vertical (black dashed) lines are the upper bounds of $\gamma_2$ set by
  the XENON 100 (right) 
  and 1T (left) experiments [Eq.~(\ref{cond-r2}) ].
 From Fig.~\ref{g2-g3} we can also see that
 there exists a parameter space with $m_\chi=135$ GeV,
 $m_{\eta_R^0} > 153$ GeV, a large $\gamma_3/4 \pi$ 
(between $0.65$ and $1.0$) and 
$\gamma_2$ satisfying the XENON constraint [Eq.~(\ref{cond-r2}) ].
A large $\gamma_3$ is needed to explain the 
$135$ GeV $\gamma$-ray line in the Fermi spectrum,
as we will see below.

\subsection{Fermi LAT $135$ GeV  $\gamma$-ray line}
\begin{figure}
  \includegraphics[width=12cm]{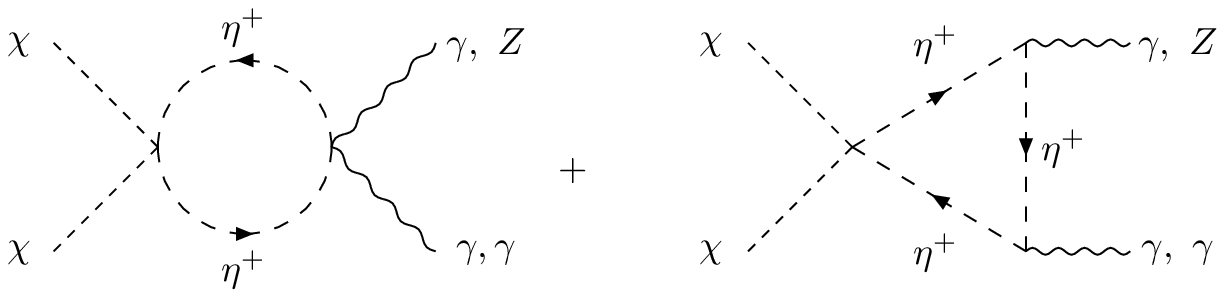}
\caption{\label{xxgg}\footnotesize
One-loop diagrams for $\chi\chi\to \gamma\gamma$.}
\end{figure}

\begin{figure}
  \includegraphics[width=13cm]{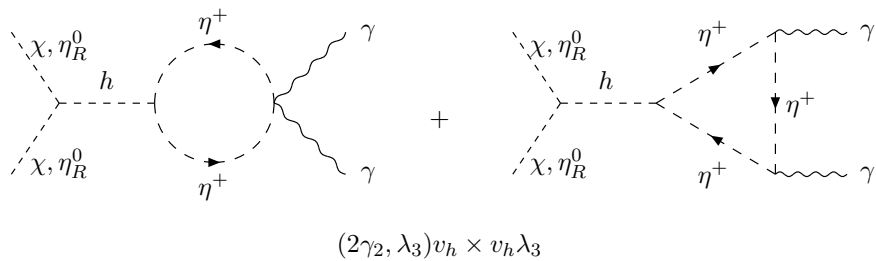}
\caption{\label{xxgg2}\footnotesize
S-channel diagrams  for $\chi\chi, \eta_R^0\eta_R^0\to \gamma\gamma$.}
\end{figure}

\begin{figure}
  \includegraphics[width=10cm]{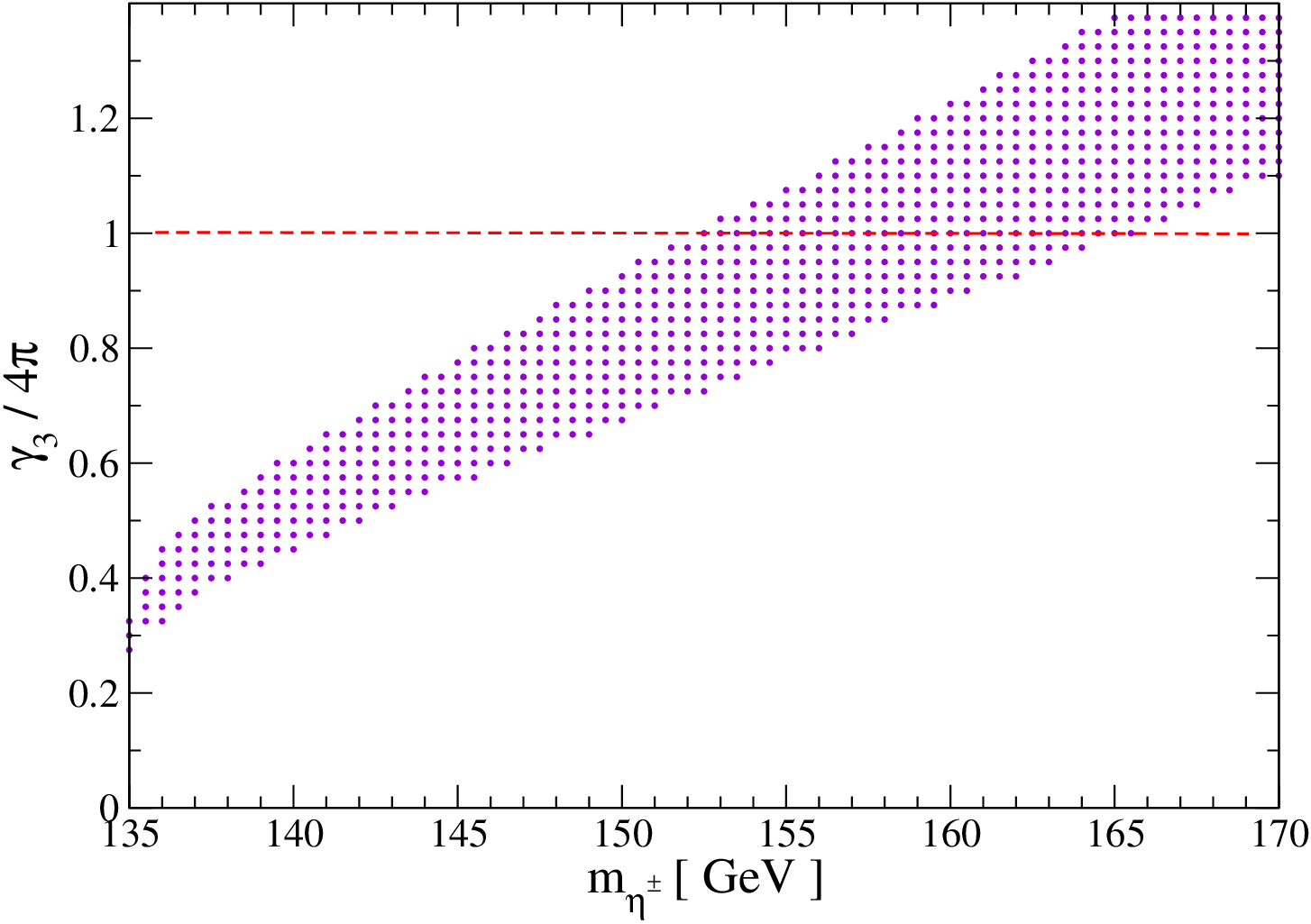}
\caption{\label{gamma3}\footnotesize
The area for $ \sigma (\chi\chi\to\gamma\gamma) v
=(1.27^{+0.37}_{-0.43})\times10 ^{-27}~
\mbox{cm}^3\mbox{s}^{-1}$ in the 
$m_{\eta^\pm}$--$\gamma_3$ plane.}
\end{figure}

\begin{figure}
  \includegraphics[width=10cm]{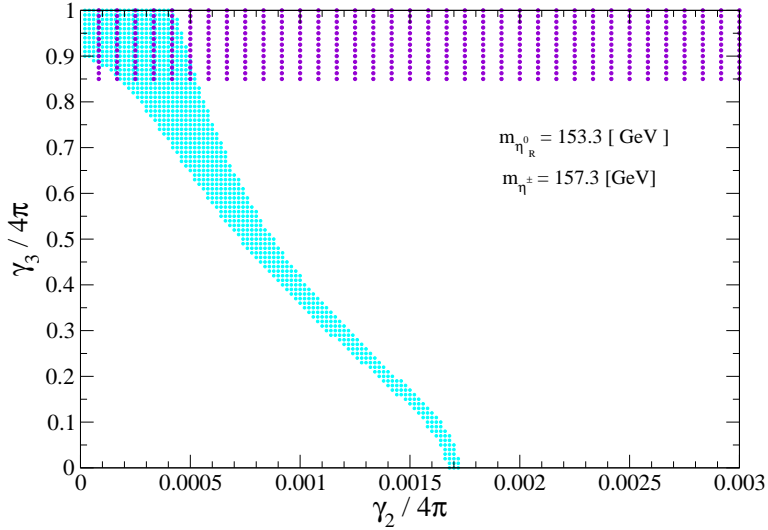}
\caption{\label{ggOmega}\footnotesize
The parameter space (overlapped area) in which 
$ \sigma (\chi\chi\to\gamma\gamma) v
=(1.27^{+0.37}_{-0.43})\times10 ^{-27}~
\mbox{cm}^3\mbox{s}^{-1}$ and $\Omega_T h^2=0.1157\pm 0.0046
~(2\sigma)$ can be obtained  for $m_{\eta_R^0}=153.3$ GeV and 
 $m_{\eta^\pm}=m_{\eta_R^0}+4~\mbox{GeV}=157.3$ GeV. }
\end{figure}

Our model contains 
the coupling between $\chi$ and $\eta^+$
(${\cal L}_{\eta^\dag \eta \chi^2}=-\gamma_3~\eta^\dag \eta \chi^2$).
Because of this coupling there are  diagrams in Fig.~\ref{xxgg},
which produce monochromatic $\gamma$ lines
through the annihilation of $\chi$.
We would like to use these diagrams to explain 
the monochromatic $\gamma$-ray line 
\cite{Bringmann:2012vr,Weniger:2012tx,Su:2012ft,Su:2012zg} 
observed at the Fermi LAT 
\cite{Atwood:2009ez,Abdo:2010nc,Ackermann:2012qk,Fermi-LAT}.
There exist also s-channel diagrams with the SM Higgs
propagator as shown in Fig.~\ref{xxgg2}. 
The $h\chi\chi$ coupling is proportional to
$\gamma_2$, while $h \eta_R^0 \eta_R^0$ coupling
 is proportional to $\lambda_3$.
The annihilation cross section 
$\sigma(\eta_R^0\eta_R^0
\to \gamma\gamma)$ is large,  because $\lambda_3$ is large
[see Eqs.~(\ref{gg}) and (\ref{cond-r3})]
\footnote{In \cite{Biswas:2013nn},
$\sigma(\eta_R^0\eta_R^0
\to \gamma\gamma)$ is used to explain the monochromatic
$\gamma$-ray line, where a fine-tuned cancellation mechanism
to suppress the total annihilation cross section
of $\eta_R^0$  \cite{LopezHonorez:2010tb} is employed }.
Besides,
due to the gauge interactions (see the right diagram of Fig.~\ref{xxww})
the relic density of $\eta_R^0$ 
is very small, 
so the annihilation of the $\eta_R^0$ DM
cannot contribute to the monochromatic $\gamma$-ray.
Furthermore, for the same reason, the tree-level 
annihilations of $\eta_R^0$ into a pair of $W$'s and $Z$'s,
which would contribute to the continuum $\gamma$-ray
spectrum, are also suppressed.
In contrast to this case, 
the pure gauge interaction (the right diagram of Fig.~\ref{xxww}) 
is absent for the annihilation of $\chi$.
The entire annihilations of $\chi$ into the SM particles
are controlled by the single coupling $\gamma_2$,
which has to satisfy the constraint of Eq.~(\ref{cond-r2}).
Therefore, we may assume that the main contribution to
$\sigma (\chi\chi\to\gamma\gamma)$ comes from the 
one-loop diagrams 
in Fig.~\ref{xxgg} and  find
\be
\sigma (\chi\chi\to\gamma\gamma) v
&=&\frac{\gamma_3^2  \alpha^2 
m_\chi^2}{32 \pi^3 m_{\eta^\pm}^4}
|F_0(m^2_{\eta^\pm}/m^2_\chi)|^2~
\nonumber \\
&\simeq &\left[\frac{\gamma_3}{4\pi}\right]^2
\left\{\begin{array}{c}
4.23  \\
2.66\\
1.87\\
1.39\\
\end{array}\right.\times10^{-27}
~\mbox{cm}^3\mbox{s}^{-1}~\mbox{for}~
m_{\eta^\pm}=\left\{\begin{array}{c}
140 \\
145\\
150\\
155\\
\end{array}\right.~\mbox{GeV}~.
\ee
These values should be compared with
$(1.27^{+0.37}_{-0.43})\times10 ^{-27}~
\mbox{cm}^3\mbox{s}^{-1}$   \cite{Weniger:2012tx} for an Einasto
 DM  galactic halo profile,
which is the size that could explain
the monochromatic $\gamma$ line 
 observed at the  Fermi LAT. 
So, if $\gamma_3/4 \pi$
is large of $O(1)$, the desired value could be obtained.
In Fig.~\ref{gamma3} we show the area 
in the $m_{\eta^\pm}$--$\gamma_3$ plane
in which
$ \sigma (\chi\chi\to\gamma\gamma) v
=(1.27^{+0.37}_{-0.43})\times10 ^{-27}~
\mbox{cm}^3\mbox{s}^{-1}$ can be obtained. 
If $\eta^\pm$ is lighter than
$\chi$, then $\chi$ can be annihilated into  a pair of $\eta^\pm$.
For $m_\chi=135$ GeV, the annihilation cross section becomes
$\sigma(\chi\chi\to\eta^+\eta^-) v
=(\gamma_3^2/2 \pi m_\chi^2)(1-m^2_{\eta^\pm}/m_\chi^2)^{1/2}
\simeq \gamma_3^2 (1-m^2_{\eta^\pm}/m_\chi^2)^{1/2}
 \times 10^{-22}~\mbox{cm}^3\mbox{s}^{-1}$,
 which is too large to obtain a sufficiently large relic density of $\chi$
 for a large $\gamma_3$ unless $m_\chi  \leq m_{\eta^\pm}$.
This is why we have to assume that $m_\chi  
 < m_{\eta^0_R}, m_{\eta^0_I} ,m_{\eta^\pm}$.
  Comparing Fig.~\ref{gamma3} with Fig.~\ref{g2-g3}, we see 
 that there is an overlapped area, that is, a parameter space
 in which $ \sigma (\chi\chi\to\gamma\gamma) v\simeq 10 \times
 10^{-27}~\mbox{cm}^3~\mbox{s}^{-1}$ and
 $\Omega_Th^2\simeq 0.12$ can be obtained. This is shown in Fig.~\ref{ggOmega}
 for $m_{\eta_R^0}=153.3$ GeV and 
 $m_{\eta^\pm}=m_{\eta_R^0}+4~\mbox{GeV}=157.3$ GeV.

The same diagrams  as in Fig.~\ref{xxgg} also produce $\gamma Z$.
The annihilation cross section is given by
\be
 \sigma(\chi\chi\to\gamma Z)v
&=&\frac{\gamma_3^2 \alpha^2 m_\chi^2
\cot^2(2\theta_W)}{32 \pi^3 m^4_{\eta^\pm}}
 |\hat{F}_0(m^2_{\eta^\pm}/m_\chi^2)|^2
(1-\frac{m_Z^2}{4m_\chi^2})~,
\ee
 where
 \be
 \hat{F}_0(\tau_\eta)
 &=& \tau_\eta \left[
 ~\frac{1}{2}-\int_0^1dx\int_0^{1-x}dy
\frac{\tau_\eta-\tau_Z(x^2+y^2)+4 xy (1-\tau_Z/2)}
{\tau_\eta+\tau_Z(x^2+y^2)-4 xy (1-\tau_Z/2)} \right]
\ee
with
$\tau_\eta=m^2_{\eta^\pm}/m^2_\chi$ and
$\tau_Z=m^2_Z/m^2_\chi$. For $m_{\eta^\pm}=
150$ GeV and $m_\chi=
135$ GeV, for instance,  we obtain $ \sigma(\chi\chi\to\gamma Z)v=
3.5 (\gamma_3/4\pi)^2 \times 10^{-29}~\mbox{cm}^3\mbox{s}^{-1}$,
which is about $2$ \% of $ \sigma(\chi\chi\to\gamma \gamma)v$. According to Ref.~\cite{Cohen:2012me}, 
this is welcome to explain the Fermi LAT monochromatic $\gamma$ line.

 We also have to satisfy the constraints on the continuum 
 $\gamma$ and follow the analyses of 
 Ref.~\cite{Cohen:2012me} (see also Refs.~\cite{Buchmuller:2012rc,Belanger:2012ta}, which give similar constraints).
  They  consider  two different constraints,
  supersaturation constraint and shape constraint,
  which can be  transferred to
the upper bound on the ratio of
the total annihilation cross section  to
     $\sigma(\chi\chi\to\gamma\gamma)$.
Specifically, 
  they consider the constraint on the theoretical ratio
    \be
  R^{\rm th}_T &=&\frac{\sigma_T}
  {2\sigma_{\gamma\gamma}+\sigma_{\gamma Z}}~,
 \ee
where  $ \sigma_T$ is the total annihilation cross section.
Since the dominant  origins for the continuum
$\gamma$ are
$\sigma(\chi\chi \to W^+W^-, ZZ, b\bar{b}, \tau^+\tau^-,
\mu^+\mu^-)$, we also consider the individual
cross sections and calculate
\be
R^{\rm th}_\alpha =\frac{\sigma(\chi\chi \to \alpha)}
  {2\sigma(\chi\chi\to \gamma\gamma)+
  \sigma(\chi\chi\to \gamma Z)}
  \simeq \frac{\sigma(\chi\chi \to \alpha)}
  {2\sigma(\chi\chi\to \gamma\gamma)}~,
  \ee
in addition to $
R^{\rm th}_{\rm SM} \simeq
\sigma(\chi\chi \to {\rm SM} )/
  2\sigma(\chi\chi\to \gamma\gamma)$, where 
  $ \alpha=W^+W^-, ZZ, f\bar{f}$.
 We use the same parameter values given in Eq.~(\ref{input3})
with $\gamma_2/4 \pi=1.0\times 10^{-4}$ and 
 $m_{\eta_R^0}=153.3$ GeV, and
we obtain
  \be
  \Omega_\eta h^2&=&0.981\times 10^{-5}~,~
    \Omega_\chi h^2=0.1197~,~
      \Omega_T h^2=0.1197~,\label{omega5}
      \ee
and
\be 
   \sigma(\chi\chi\to \gamma\gamma) v      
   &\simeq & 
1.2\times 10^{-27}~\mbox{cm}^3\mbox{s}^{-1}~,\label{gg2}\\
 \sigma(\chi\chi\to \mbox{SM}) v &\simeq & 
8.0\times 10^{-29}~\mbox{cm}^3\mbox{s}^{-1}
 ~,~R_{\rm SM}^{\rm th} \simeq 0.06~, \label{rth1}\\
 \sigma(\chi\chi\to W^+W^-) v &\simeq & 
3.9\times 10^{-29}~\mbox{cm}^3\mbox{s}^{-1}
~,~R_W^{\rm th} \simeq 0.03~,\label{rth2} \\
 \sigma(\chi\chi\to Z Z) v &\simeq &  
1.7\times 10^{-29}~\mbox{cm}^3\mbox{s}^{-1}
~,~ R_Z^{\rm th} \simeq 0.01~,\label{rth3}\\
\sigma(\chi\chi\to h h) v &\simeq &  
2.5\times 10^{-29}~\mbox{cm}^3\mbox{s}^{-1}
~,~ R_h^{\rm th} \simeq 0.02~,\label{rth4}\\
\sigma(\chi\chi\to f \bar{f}) v &\simeq &  
1.1\times 10^{-31}~\mbox{cm}^3\mbox{s}^{-1}~
~,~ R_f^{\rm th} \simeq 10^{-4}~.\label{rth5}
   \ee
  These values of $R^{\rm th}$ should satisfy the 
     supersaturation constraint as well as the shape constraint
     of  Ref.~\cite{Cohen:2012me} (see also Ref.~\cite{Buchmuller:2012rc}).
There are also constraints coming from
the antiproton-to-proton flux observed by 
the PAMELA \cite{Adriani:2010rc}.
Antiprotons can be produced by the DM annihilations into
the gauge bosons, Higgs bosons and quarks.
To explain the PAMELA data, these productions
have to be suppressed.  The annihilation cross sections $\times v$ given in
 Eqs.~(\ref{rth1})--(\ref{rth5})  satisfy all the constraints, 
including the most stringent one,
$\lsim 10^{-26}~\mbox{cm}^3\mbox{s}^{-1}$
\cite{
Cohen:2012me,Buchmuller:2012rc,Belanger:2012ta,
Ibarra:2008qg,Donato:2008jk,Buchmuller:2009xv,
Asano:2012zv}.
So, the model could
 explain the monochromatic $\gamma$ line 
 observed at the Fermi LAT if $m_{\eta^\pm}
 \simeq 153$ GeV and $\gamma_3/4 \pi
\sim O(1)$, which is at the border of perturbation theory.

Therefore, in  this parameter space
the model has a meaning only up to a scale slightly above the electroweak
scale, and the scalar potential is stable only below that energy.
In this sense, the SM is more natural in the parameter space
in which the model has a potential to explain  the
$\gamma$ excess   both  in the Higgs decay and at the Fermi LAT.
In other parameter space, where the scalar couplings
are small
but the  $\gamma$ excesses cannot be 
(or do not have to be) simultaneously explained, the model
can remain well defined to a very high energy close to the Planck scale.
This situation would occur if the monochromatic $\gamma$ ray 
in the Fermi data turn out to be an instrumental effect 
(see for instance Refs.~\cite{Whiteson:2013cs,Weniger:2013dya}).
Then $\gamma_3$ does not have to be large, and the $\chi$ DM does not have 
to be lighter than the $\eta$ DM, so that, to explain the diphoton
mode in the Higgs decay, the scalar coupling $\lambda_3$ 
can become much smaller (because the decay mode 
can be enhanced by
a smaller mass of the charged $\eta^\pm$
\cite{Arhrib:2012ia,Swiezewska:2012eh}).
In this parameter space, the model can remain 
perturbative for a wide range of energy scales.

\section{Conclusion and discussion}
In this paper we have proposed 
a non-supersymmetric model of a two-loop radiative seesaw,
in which the lepton number is softly broken
by a dimension-two operator, and
the tree-level Dirac mass is forbidden by $Z_2 \times Z_2'$.
This discrete symmetry can be used to stabilize 
 two or three dark matter particles.
The model contains, in addition to the SM Higgs field, 
an inert $SU(2)_L$ doublet scalar $\eta$ and two inert 
singlet scalars $\phi$ and $\chi$, and this is a minimal set.
We have considered the SM Higgs boson decay into two $\gamma$'s 
and found that it is enhanced 
by $\eta^+$ circulating 
in   one-loop diagrams for $h\to\gamma\gamma$.
 $\eta^+$ is also circulating in  similar one-loop diagrams
contributing to $\chi\chi\to \gamma\gamma$, and
we have found that the model has a potential to
explain the Fermi LAT $135$ GeV $\gamma$-ray  line.

The mechanism to explain the Fermi LAT $135$ GeV 
$\gamma$-ray line
in the present model is strongly based on the fact that 
there exist some particles of similar masses where at least one of them is DM.
Let us briefly outline the mechanism \cite{Griest:1990kh,Tulin:2012uq}.
Annihilation (or decay) of DM into $\gamma$'s happens always at the loop  level.
Those into the SM particles, i.e.
$W$, $Z$, Higgs boson pairs, etc.,
are usually  possible  at the tree level, and they produce
continuum $\gamma$ rays as well as antiprotons.
To explain  the  $135$ GeV $\gamma$-ray line, we have to
suppress these tree-level processes somehow, or enhance
the loop process, while keeping the  relic abundance of DM
at the observed value.
In the present  model this is realized in the following way:
There are two kinds of the tree-level  DM annihilations;
one into the  SM particles and the other into
a pair of other DM particles (DM conversion).
The slightly heavier DM ($\eta_R^0$ in our model) has 
large annihilation cross sections both 
into the  SM particles and other DM particles, 
so its relic density is very small.
The  $135$ GeV $\gamma$-ray line comes mainly from
the annihilation of the slightly lighter DM ($\chi$ in our model).
Its annihilation cross section  into the  SM particles 
has to be sufficiently small  to  suppress the  continuum
$\gamma$ rays and the production of antiprotons.
Although the annihilation of the lighter DM into
heavier DM is kinematically forbidden at zero temperature,
this process becomes operative at high temperature:
The smaller the mass difference of two DM particles is, the more effective
is the DM conversion.
So, at high temperature 
the annihilation of the lighter DM  is  controlled by
the mass deference and can be large, but at low temperature,
this conversion process practically disappears (see Fig.~\ref{sigmaeff}).
(For this mechanism to work, the slightly heavier particle 
does not have to be stable. )
It is, however, important that the lighter one be SM gauge singlet
to avoid the fact that the tree-level annihilations are entirely controlled
by the SM gauge interactions.
(In the present model, the $\chi$ DM is not {\it ad hoc} introduced.)
This is the reason why we can obtain the observed 
value of the relic density for the lighter DM,
although the annihilation cross sections
of the lighter DM into the SM particles are very small 
$\lsim O(10^{-27})~\mbox{cm}^3\mbox{s}^{-1}$
in the galaxy.

The annihilation cross sections into the SM particles given in 
Eqs.~(\ref{omega5}), (\ref{rth1})--(\ref{rth5}) are obtained
without one-loop corrections.
Strictly speaking, we should include the one-loop corrections,
because  the tree-level contributions are so small that 
the one-loop corrections
may be larger than the tree-level corrections.
Similarly, the relic densities for Fig.~
\ref{g2-g3} and also Eq.~(\ref{omega5}) have been computed
by neglecting the coannihilation of the DM particles with
the  charged and  CP-odd
components of $\eta$, although we have assumed that 
their mass differences are not large [ i.e.
$m_{\eta^\pm} -m_\chi=(17-25)~\mbox{ GeV}$, 
$m_{\eta^\pm} -m_{\eta_R^0}=4~\mbox{ GeV}$ 
and $m_{\eta^\pm}=m_{\eta_I^0}$ ].
The one-loop corrections would change $\gamma_2$ effectively.
That is, one-loop corrections can be partially absorbed into 
$\gamma_2$, so that 
the annihilation cross sections [ Eq.~(\ref{rth1})--(\ref{rth5}) ] would change
only slightly;
to  transgress the supersaturation constraint and shape constraint,
a change of 2 orders of magnitude is needed.
  The coannihilations also would  effectively 
  increase  $\gamma_3$.
  To obtain a realistic relic abundance for the $\chi$ DM
in this situation,
 the mass of the $\eta$ DM
  mass should be slightly increased, as one can see from Fig.~\ref{g2-g3}.
The one-loop analysis including the coannihilations
is beyond the scope of the present paper,
and we will leave it for our future project.

We have assumed throughout that the $\phi$ is so heavy that
it decays into a $\eta_R^0$ and a $\chi$.
In the case that $m_{\phi_R} < m_{\chi}+m_{\eta_R^0}$, 
it becomes the third DM, whose annihilation
may be responsible for the second monochromatic $\gamma$-ray line
in the Fermi data \cite{Rajaraman:2012db}. 
We leave this question to the future program.

\vspace*{5mm}
We thank Michael Duerr, Maria Krawczyk and Abdesslam Arhrib
for useful discussions.
 M.~A. thanks the Max-Planck-Institut f\"ur Kernphysik,
Heidelberg for kind hospitality.
The work of M.~A.\ is supported in part by the Grant-in-Aid for Scientific 
Research for Young Scientists (B) (Grant No. 22740137), 
and J.~K.\ is partially supported by the Grant-in-Aid for Scientific
Research (C) from the Japan Society for Promotion of Science (Grant No. 22540271).

\end{document}